\documentclass[twoside,11pt]{article}
\usepackage{epsfig}
\usepackage{amssymb,amsbsy,amsmath,amsfonts}
\def\beq{\begin{equation}}
\def\eeq{\end{equation}}
\def\bea{\begin{eqnarray}}
\def\eea{\end{eqnarray}}
\def\beqa{\begin{equation}\begin{array}{l}}
\def\eeqa{\end{array}\end{equation}}
\def\eqlab#1{\label{eq:#1}}
\def\figlab#1{\label{fig:#1}}
\def\tablab#1{\label{tab:#1}}

\def\barr{\left(\begin{array}{c}}
\def\earr{\end{array}\right)}
\def\bmat{\left(\begin{array}{cc}}
\def\emat{\end{array}\right)}
\def\Figref#1{Fig.~\ref{fig:#1}}

\def\Tabref#1{Table \ref{tab:#1}}

\def\al{\alpha}

\def\ga{\gamma} 
\def\de{\delta} \def\De{\Delta}
\def\vDe{\varDelta}
  \def\eps{\epsilon}

\def\la{\lambda} \def\La{{\Lambda}}

\def\si{\sigma} 
  
\def\w{\omega}

\def\pa{\partial}

\def\pa{\partial}

\def\lag{{\mathcal L}}

\def\mathscr{\mathcal}

\def\3d{3-D}

\def\ol#1{\overline{#1}}

\def\ceft{$\chi$EFT}

\begin{document}

\title{\vspace{1cm} The $\Delta$(1232) Resonance in Chiral Effective Field Theory\thanks{Based on a seminar given at Erice School
``Quarks in Hadrons and Nuclei'', 29th Course, 16--24 Sep 2007, Sicily.
Supported  by the European Community-Research Infrastructure Activity under FP6 contract RII3-CT-2004-506078.}
%\thanks{Based on a seminar given at Erice School
%``Quarks in Hadrons and Nuclei'', 29th Course, 16--24 Sep 2007, Erice, Sicily,
%Italy}
}
\author{Vladimir Pascalutsa\\
\\
\it ECT* Trento,  Villa Tambosi, Villazzano (TN),  I-38050,  Italy}

\date{}
\maketitle

\begin{abstract}
I discuss the problem of formulating the baryon chiral perturbation theory ($\chi$PT) in the presence of a light resonance, such as the $\Delta(1232)$, the lightest nucleon resonance. It is shown how to extend the power counting of $\chi$PT to correctly account for the resonant contributions. Recent applications of the resulting chiral effective-field
theory to the  description of pion production reactions in $\Delta$-resonance region are briefly reviewed.
\end{abstract}

\section{Introduction}
\label{sec:sec1}

A quantitative description of the low-energy physics of nucleons and pions based on the underlying theory of the strong interaction, quantum chromodynamics (QCD), is still lacking, and the Millennium prize for a solution of this problem
is still outstanding. Lattice QCD holds a promise to solve the problem one day by a sheer brute force, through a Monte-Carlo simulation of QCD in discrete Euclidean spacetime. In the absence of such a solution, the most appropriate framework for the description of the low-energy strong interaction is Chiral Perturbation Theory ($\chi$PT), an effective field theory of QCD written directly in terms of the hadronic degrees of freedom~\cite{Weinberg:1978kz}.
\newline
\indent
The main guiding principle in the construction
of $\chi$PT is the chiral symmetry of massless QCD Lagrangian and the pattern
of its breaking, which allows to organize the effective Lagrangian
in powers of derivatives of the Goldstone boson fields -- the pions,  schematically:
\beq
{\cal L}(\pi, N, \ldots) = \sum_{n} {\mathcal L}^{(n)}=
\sum_{n} {\cal O}_n(c_i)
\frac{(\pa \pi)^n}{\La^n}
\eeq
where
$O_n$ are some field operators which may contain pion fields
but not their derivatives. The all-possible field operators,
constrained by chiral and other
symmetries, appear with the free parameters, $c_i$,
the so-called low energy constants (LECs).
The mass scale $\La$ is the heavy scale
which sets the upper limit of applicability of $\chi$PT
and is believed to be of order of $1$ GeV, the scale
of spontaneous chiral symmetry breaking that leads to
the appearance of the Goldstone bosons.
\newline
\indent
 This expansion
of the Lagrangian translates into a low-energy expansion
of the $S$-matrix:
\beq
S = \sum_{n}\, A_n( c_i) \, \frac{ p^n}{\La^n}
\eeq
where $A$'s are amplitudes which depend on LECs, and $p$ denotes the
typical momentum of the particles. 
If an analogous expansion could be obtained
directly from QCD, it would be equivalent to the $\chi$PT one,
provides the LECs are matched to the QCD coupling,
i.e., $c_n = c_n (\La_{QCD})$. At present however
the best one can do is to match the LECs to experimental data, and hope they take reasonable (natural) values, such that
this above expansion is convergent. 
\newline
\indent
One case where the convergence of the $\chi$PT expansion is
immediately questioned is the case of hadronic bound states and
resonances. In the presence of a bound state or a resonance
the low-energy expansion of the $S$-matrix
goes as: 
\beq
S \sim \sum_{n} A_n \, \left(
\frac{p}{ { \Delta E} }\right)^n,
\eeq
where ${\Delta E}$ is the excitation (binding) energy of
the resonance (bound state). Thus, the limit
of applicability of $\chi$PT is limited not by $\La\sim 1$ GeV but by
the characteristic energy scale $\De E$ of the closest bound
or excited state.
Furthermore, in the vicinity of a bound state or a resonance
the $S$-matrix has a pole, which cannot be reproduced in a
purely perturbative expansion in energy that is
utilized in $\chi$PT.
\newline
\indent
This problem arises in various contexts, ranging from
pion-pion scattering\cite{Caprini:2005zr}
to halo nuclei\cite{vanKolck:2004te}.
Here I shall consider the case of the $\De(1232)$, which
is an ideal study case for the problem of resonances in $\chi$PT.
It is relatively light, with the excitation energy of $\vDe\equiv M_\De -M_N
\approx 300$ MeV, elastic, and well separated from the other
nucleon resonances. It is also a very prominent resonance
and plays an important
role in many processes, including astrophysical ones, e.g., it is
responsible for the the so-called GZK cutoff
(damping of the high-energy cosmic rays by the
cosmic microwave background).

\section{Power counting for the $\De$ resonance}

Let me start with a simple example: Compton scattering on the nucleon.
The total cross-section of this process, as the function
of photon energy $\w$, is shown in
\Figref{totalcs}. In this case we are
able to examine the entire energy range,
starting with the soft-photon limit $\w\simeq 0$, through the pion production
threshold $\w\simeq m_\pi$ and into the resonance region
$\w\sim \vDe$.
\begin{figure}[b]
\centerline{  \epsfxsize=8cm
  \epsffile{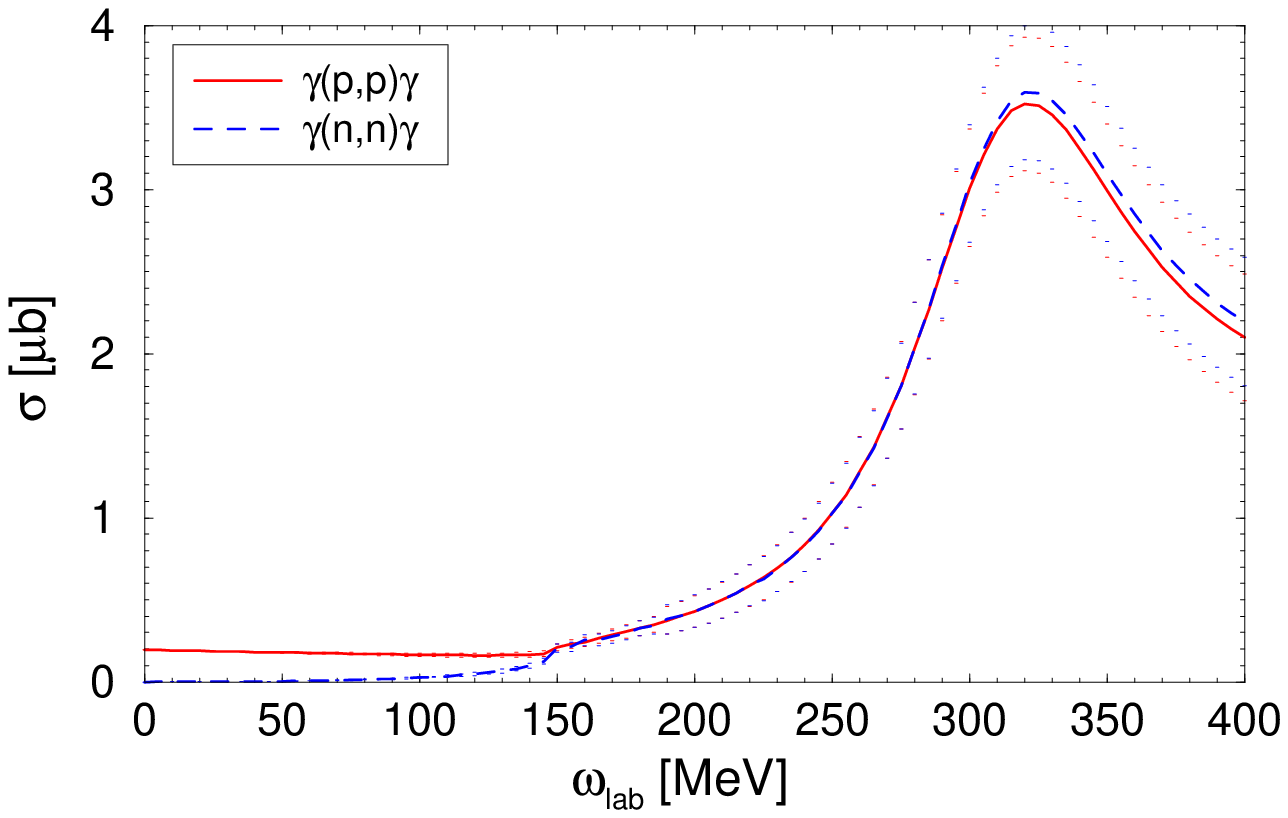}
}
\caption{(Color online)
Total cross-section of the Compton scattering on the
nucleon (proton -- red solid curve, neutron -- blue dashed curve),
as the function of the incident photon lab energy. The curves
are obtained in a $\chi$EFT calculation\cite{Pascalutsa:2002pi}.}
\figlab{totalcs}
\end{figure}
\newline
\indent
At energies up to around the pion production
threshold the cross section shows a smooth behavior
which can reproduced by a low energy expansion.
In this region the $\De$-resonance
 can be ``integrated out'', as its tail contribution can be
mimicked by the terms already present in the $\chi$PT Lagrangian
with nucleons only\cite{GSS89}. 
\newline
\indent
Higher in energy, however, the rapid energy variation induced
by the resonance pole is not reproducible by a naive low-energy
expansion. Obviously, to describe this behavior it is necessary to
introduce the $\De$ as an explicit degree of freedom, hence
include a corresponding field in the effective chiral Lagrangian.
The details of how this is done have recently been reviewed
in\cite{Pascalutsa:2006up}.
\newline
\indent
Once the $\De$ appears in the Lagrangian the question is how
to power-count its contributions. In $\chi$EFT with
pions and nucleons alone the power-counting index of a graph
with $L$ loops, $N_\pi$ ($N_N$) internal pion (nucleon)
lines, and $V_k$ vertices from $k$th-order Lagrangian
is found as
\beq
\eqlab{chptindex}
n_{\chi \mathrm{PT}} =  4 L  - 2 N_\pi - N_N + \sum_k k V_k\,.
\eeq
What about the graphs with the $\De$, such as those depicted
in \Figref{ODR} ? Their power counting turns out to be
dependent on how one weighs the excitation energy $\vDe$
in comparison with the other mass scales of the theory.
In this case we have the soft momentum $p$ (or, $\w$),
the pion mass $m_\pi$, and heavy scales which we collectively
denote as $\La$.
\newline
\indent
The Small Scale Expansion\cite{HHK97} (SSE)
counts all light scales equally: $p\sim m_\pi \sim \vDe$.
The small parameter is then:
$
\eps = \left\{p/\La,\, m_\pi/\La,\, \vDe/\La\right\}\,.
$
An unsatisfactory feature of such a  counting
is that the $\De$-resonance
contributions are always estimated to be of the
same size as the nucleon contributions.
As we have seen from \Figref{totalcs}, in reality
the resonance contributions
are suppressed at low energies while being  dominant
in the resonance region.
Therefore, the power counting
{\it overestimates} the $\De$-contributions
at lower energies and {\it underestimates} them
at the resonance energies.
\begin{figure}[t,b]
\centerline{  \epsfxsize=8cm
  \epsffile{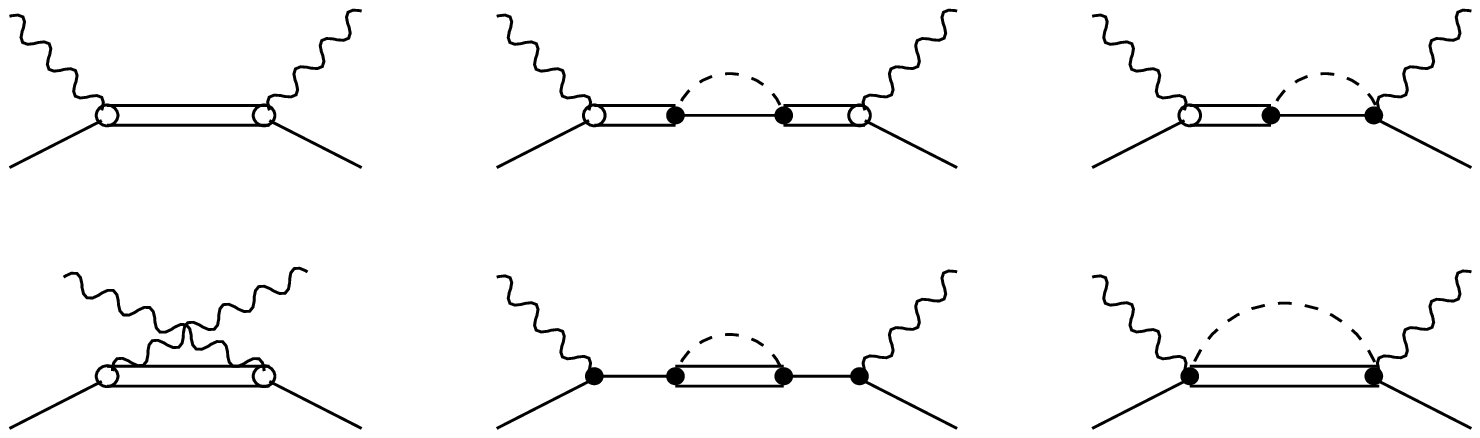}
}
\caption{Examples of the one-Delta-reducible (1st row) and
 the one-Delta-irreducible (2nd row) graphs in Compton scattering. }
\figlab{ODR}
\end{figure}
\newline
\indent
A more adequate power counting is achieved by separating out
the resonance energy, e.g., maintaining the following
scale hierarchy, $m_\pi\ll \vDe\ll \La$,
within the power-counting scheme.
In the so-called
{\it ``$\de$ expansion''}\cite{Pascalutsa:2002pi}
this is done by introducing a small parameter $\de=\vDe/\La$,
and then counting $m_\pi/\La$ as $\de^2$. The power 2 is chosen
here because it is the closest integer representing the ratio
of these scales in the real world.
\newline
\indent
Obviously, the power
counting of the $\De$ contributions then becomes dependent on the
energy domain: in the {\it low-energy region} ($p\sim m_\pi$)
and the {\it resonance region} ($p\sim \vDe$), the momentum
counts differently.
This dependence
most significantly affects the counting of the
one-Delta-reducible (ODR) graphs. The 1st row of graphs
in \Figref{ODR}
illustrates examples of the ODR graphs for the Compton scattering
case. These graphs are all characterized by having
a number of ODR propagators, each going as
\beq
S_{ODR}\sim \frac{1}{s-M_\De^2} \sim \frac{1}{2M_\De}\frac{1}{p-\vDe}\, ,
\eeq
where $s=M_N^2 + 2M_N\w$ is the Mandelstam variable, and
the soft momentum $p$ in this case given by the photon energy.
In contrast, the nucleon propagator in analogous graphs would go
simply as $S_N\sim 1/p$.
Therefore, in the low-energy region, the $\De$ and nucleon
propagators would
count respectively as ${\mathcal O}(1/\de)$ and ${\mathcal O}(1/\de^2)$,
the $\De$ being suppressed by one power of the small parameter
as compared to the nucleon.
In the resonance region, the ODR graphs obviously
all become large. Fortunately, they all can be subsumed, leading
to ``dressed'' ODR graphs with a definite power-counting index.
Namely, it is not difficult to see that the resummation of
the classes of ODR graphs results
in ODR graphs with only a single ODR propagator of
the form
\beq
S_{ODR}^\ast = \frac{1}{S_{ODR}^{-1} - \Sigma }
\sim \frac{1}{p-\vDe-\Sigma}\,,
\eeq
where $\Sigma$ is the $\De$ self-energy.
The expansion of the self-energy begins with $p^3$, and hence
in the low-energy region
does not affect the counting of the $\De$ contributions. However,
in the resonance region the self-energy not only ameliorates
the divergence of the ODR propagator at $s=M_\De^2$ but also
determines power-counting index of the propagator.
Defining the $\De$-resonance region formally as the region of $p$
where
\beq
|p-\vDe | \leq \de^3 \La\,,
\eeq
we deduce that an ODR propagator, in this region, counts
as ${\mathcal O}(1/\de^3)$. Note that the nucleon propagator in
this region counts as ${\mathcal O}(1/\de)$, hence is
suppressed by two powers as compared to ODR propagators.
Thus, within the power-counting scheme we have the mechanism for
estimating correctly  the relative size
of the nucleon and $\De$ contributions in the two energy domains.
In \Tabref{counting} we summarize the counting of the nucleon,
ODR, and one-Delta-irreducible (ODI) propagators in
both the $\eps$- and $\de$-expansion.
\begin{table}[t]
{\centering \begin{tabular}{||c|c||c|c||}
\hline
&  $\eps$-expansion &
\multicolumn{2}{|c||}{$\de$-expansion} \\
\cline{2-4}
  & $p/\La_{\chi SB}\sim \eps$  &  $p\sim m_\pi$ & $p\sim \vDe$ \\
\hline
$\,S_N\,$ &  $1/\eps$ & $1/\de^2$ & $1/\de$\\
$\,S_{ODR}\,$ & $1/\eps$ & $1/\de$ & $1/\de^3$\\
$\,S_{ODI}\,$ & $1/\eps$ & $1/\de$ & $1/\de $ \\
\hline
\end{tabular} \par }
\caption{The counting for the nucleon, one-Delta-reducible (ODR), and
one-Delta-irreducible (ODI) propagators in the two different expansion
schemes. The counting in the $\de$-expansion depends on the energy domain.}
\tablab{counting}
\end{table}
\newline
\indent
In the following I will show two applications of the $\de$ expansion
to the calculation of processes in the $\De$-resonance region.

\section{Pion electroproduction}

The pion electroproduction on the proton in the $\De$-resonance
region has been under an intense study at many electron beam facilities,
most notably at MIT-Bates, MAMI, and Jefferson Lab. The primary
goal of these recent experiments is to measure
electromagnetic $N\to\Delta$ transition, which comes in three different
multipoles: $M1$, $E2$, and $C2$. On the theory side, these
form factors have been studied in both the SSE\cite{Gellas:1998wx,Gail}
and the $\de$-expansion\cite{Pascalutsa:2005ts,Pascalutsa:2005vq}.
\newline
\indent
The $N \to \De$ transition can be induced by a pion or a photon.
The corresponding effective Lagrangians are written as:
\begin{subequations}
\eqlab{lagran}
\bea
\lag^{(1)}_{N\De} &=&  \frac{i h_A}{2 f_\pi M_\De}
\ol N\, T^a \,\ga^{\mu\nu\la}\, (\pa_\mu \De_\nu)\, \pa_\la \pi^a
+ \mbox{H.c.}, \\
\lag^{(2)}_{N\De} &=&   \frac{3 i e g_M}{2M_N (M_N + M_\Delta)}\,\ol N\, T^3
\,(\pa_{\mu}\De_\nu) \, \tilde F^{\mu\nu}  + \mbox{H.c.},\\
\lag^{(3)}_{N\De} &=&  \frac{-3 e}{2M_N (M_N + M_\Delta)} \ol N \, T^3
\ga_5 \left[ g_E (\pa_{\mu}\De_\nu)
+  \frac{i g_C}{M_\De} \ga^\al 
(\pa_{\al}\De_\nu-\pa_\nu\De_\al) \,\pa_\mu\right] F^{\mu\nu}+ \mbox{H.c.},
\;\;\;\;\;
\eea
\end{subequations}
where $N$, $\De_\mu$, $\pi$ stand respectively for the
nucleon (spinor, isodublet), $\Delta$-isobar (vector-spinor, isoquartet),
pion (pseudoscalar, isovector) fields;  $F^{\mu\nu}$ and $\tilde F^{\mu\nu}$
are the electromagnetic field strength and its dual,
$T^a$ are the isospin-1/2-to-3/2 transition $(2\times 4$) matrices.
The coupling constants $h_A$, $g_M$, $g_E$, and $g_C$ are
the LECs describing the $N\to\De$ transition at the tree level.
\newline
\indent
We consider now the pion electroproduction on the nucleon
 to NLO in the $\de$ expansion.
Since
we are using the one-photon-exchange approximation,\footnote{For first analyses
of the two-photon-exchange effects in the $\ga N\to \De$
transition see Refs.~\cite{Pascalutsa:2005es,Kondratyuk:2006ig}.}
the pion photoproduction can be viewed as
the particular case of electroproduction at $Q^2=0$.
\begin{figure}[t]
\centerline{  \epsfxsize=11cm
  \epsffile{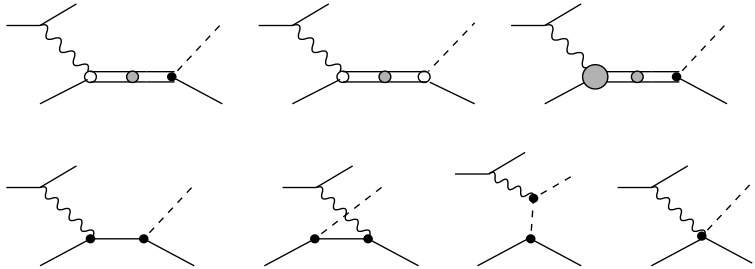}
}
\caption{Diagrams for the $e N \to e \pi N $ reaction
to LO and  NLO in the $\delta$-expansion. The dots denote
the vertices from the 1st-order Lagrangian, while the
circles are the vertices from the 2nd order Lagrangian ({\it e.g.}, the
$\ga N\De$-vertex in the first two graphs is the $g_M$ coupling from $\lag^{(2)}$).}
\figlab{diagrams}
\end{figure}
\begin{figure}[t]
\centerline{  \epsfxsize=9.5cm
  \epsffile{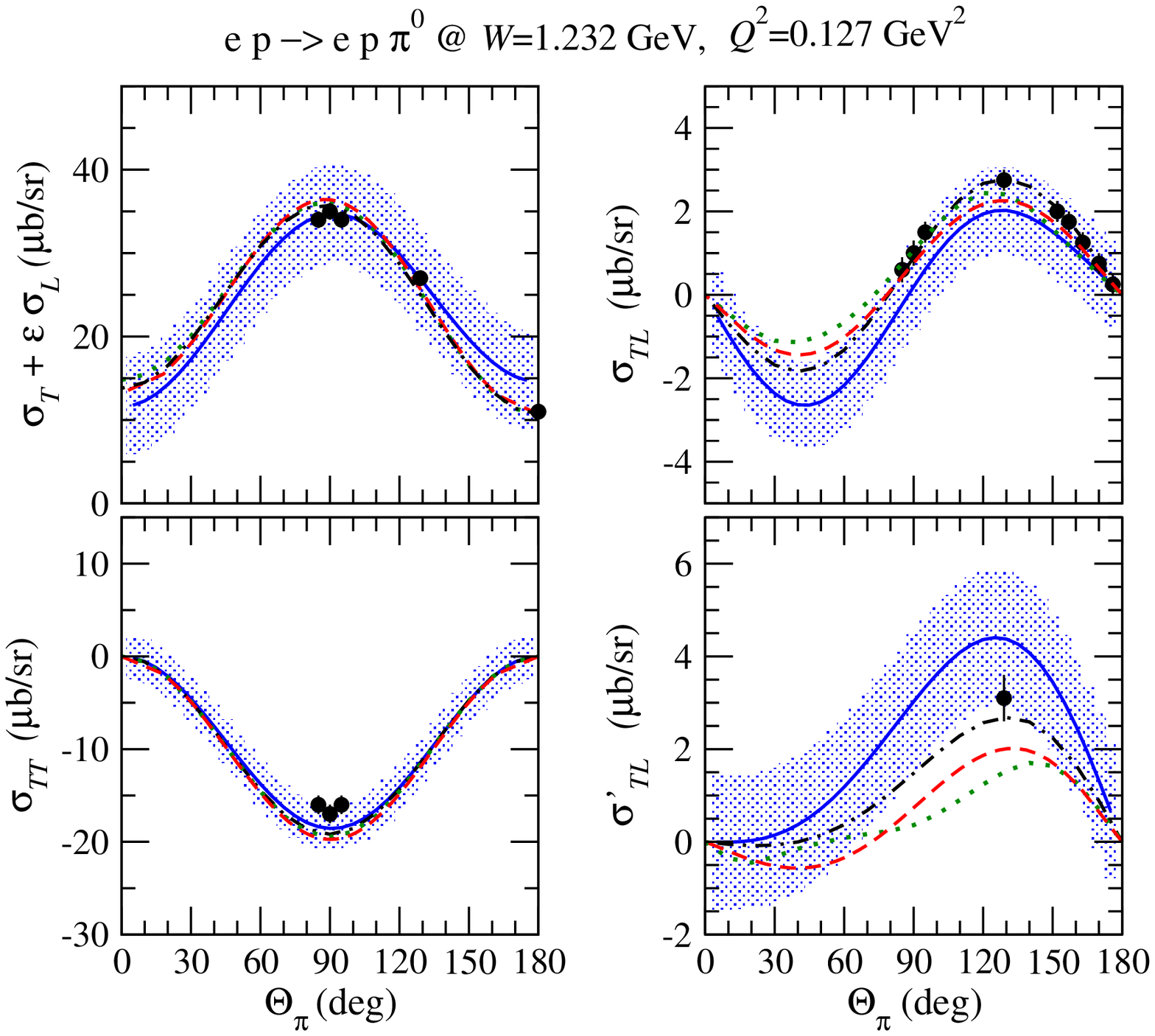}
}
\caption{The pion angular dependence
of the $\ga^\ast p \to \pi^0 p$ cross sections at
$W = 1.232$~GeV and $Q^2$ = 0.127~GeV$^2$.
Dashed-dotted (black) curves: DMT model~\cite{KY99}.
Dashed (red) curves: SL model~\cite{SL}.
Dotted (green) curves : DUO model~\cite{DUO}.
Solid (blue) curves:
\ceft \ results \cite{Pascalutsa:2005ts,Pascalutsa:2005vq}.
The bands provide an estimate of the theoretical error for the \ceft \
calculations.
Data points are from BATES
experiments~\cite{Mertz:1999hp,Kunz:2003we,Sparveris:2004jn}.
}
\label{fig:epio_cross2}
\end{figure}
The pion electroproduction amplitude to NLO in the $\de$-expansion, in the
resonance region, is given by the graphs in  \Figref{diagrams},
where the shaded
blob in the 3rd graph denotes the NLO $\ga N\De$ vertex.
The 1st graph in \Figref{diagrams} enters at the LO,
which here is ${\mathcal O}(\de^{-1})$.
All the other graphs in \Figref{diagrams}
are of NLO$={\mathcal O}(\de^{0})$.
\newline
\indent
%
%\newpage
In Fig.~\ref{fig:epio_cross2},
the different virtual photon absorption cross sections
around the resonance position are displayed
at $Q^2 =0.127$~GeV$^2$,
where recent precision data are available.
We compare these data with the present
\ceft \ calculations as well as with the results of
SL, DMT, and DUO models~\cite{SL,KY99,DUO}.
In the \ceft\ calculations, the low-energy constants
$g_M$ and $g_E$, were fixed from the resonant pion photoproduction
multipoles. Therefore, the only other low-energy constant from the
chiral Lagrangian entering the NLO calculation is $g_C$. The main
sensitivity on $g_C$ enters in $\sigma_{TL}$. A best description
of the $\si_{TL}$ data at low $Q^2$ is obtained by choosing $g_C =
-2.6$.
One sees that
the NLO \ceft\ calculation, within its accuracy, is consistent
with the experimental data for these observables at low $Q^2$.
\newline
\indent
Since the low-energy constants $g_M$, $g_E$, and $g_C$ are fixed to experiment,one can provide a prediction for the $m_\pi$ dependence of the
$\gamma N \Delta$ transition. The study of the $m_\pi$-dependence
is crucial to connect to lattice QCD results, which at present
can only be obtained for
larger pion masses.
\begin{figure}[t,b]
\centerline{  \epsfxsize=7cm%
  \epsffile{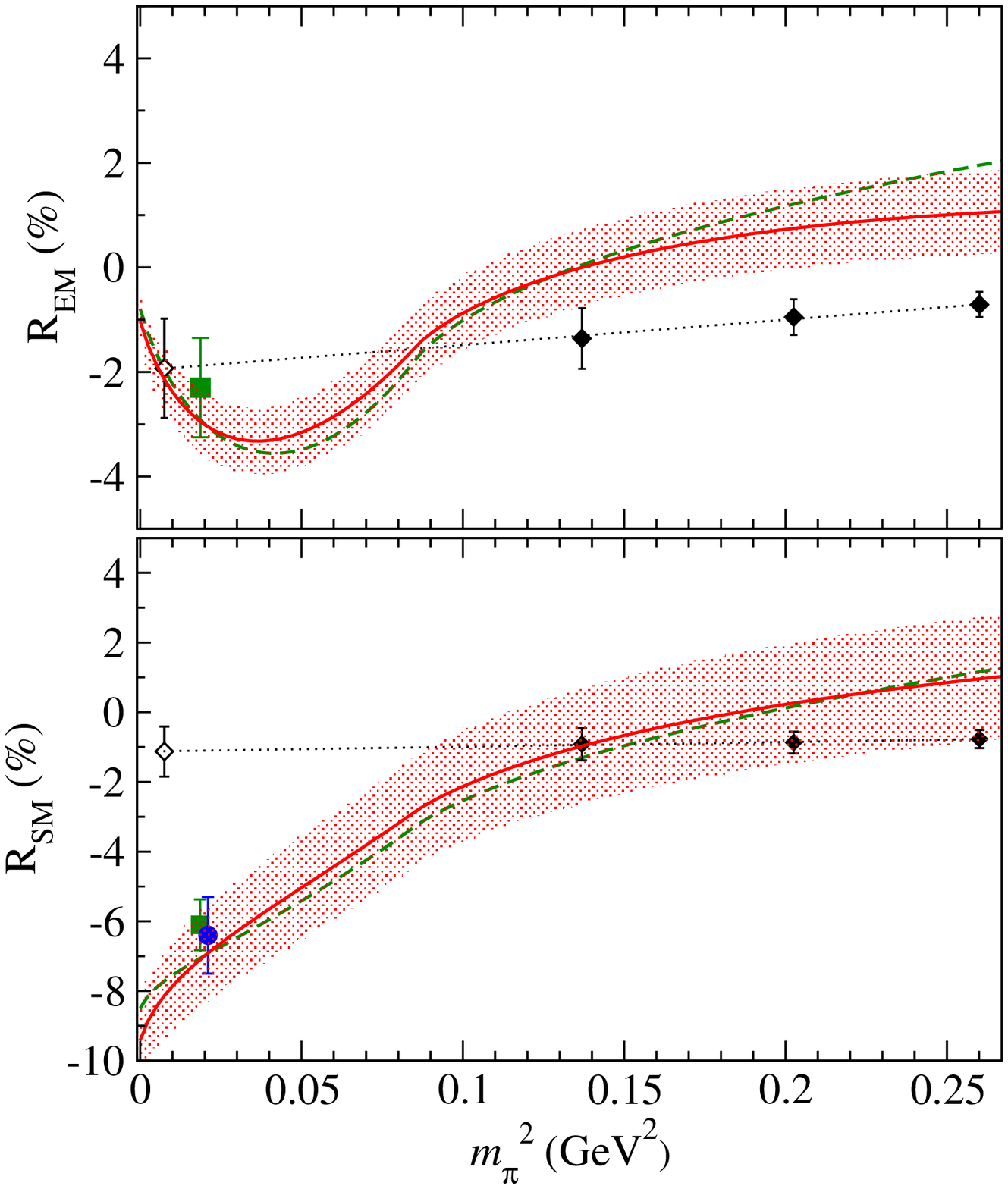}
}
\caption{
The pion mass dependence of
 $R_{EM}$ (upper panel) and
$R_{SM}$ (lower panel), at $Q^2=0.1$ GeV$^2$.
The blue circle is a data point from MAMI~\protect\cite{Pospischil:2000ad},
the green squares are data points from
BATES~\protect\cite{Mertz:1999hp,Sparveris:2004jn}.
The three filled black diamonds at larger $m_\pi$  
are lattice calculations~\protect\cite{Alexandrou:2004xn},
whereas the open diamond near $m_\pi \simeq 0$ 
represents their extrapolation assuming linear dependence in $m_\pi^2$.
Red solid curves: NLO result when accounting for the $m_\pi$ dependence in
$M_N$ and $M_\Delta$;
green dashed curves: NLO
result of Ref.~\cite{Pascalutsa:2005ts}, where  
the $m_\pi$-dependence of $M_N$ and $M_\Delta$ was not accounted for.
The error bands represent the estimate of theoretical uncertainty for
the NLO calculation.
}
\figlab{ratios}
\end{figure}
In \Figref{ratios}, one sees the $m_\pi$-dependence of the ratios
$R_{EM}=E2/M1$ and $R_{SM}=C2/M1$ and compare them to lattice QCD calculations.  The recent state-of-the-art lattice calculations of
$R_{EM}$ and $R_{SM}$~\cite{Alexandrou:2004xn} use a {\it linear},
in the quark mass ($m_q\propto m_\pi^2$), extrapolation
to the physical point, 
thus assuming that the non-analytic $m_q$-dependencies are  negligible.
The thus obtained value for $R_{SM}$ at the physical
$m_\pi$ value displays a large
discrepancy with the  experimental result, as seen in \Figref{ratios}.
This \ceft\ calculation,
on the other hand, shows  that the non-analytic dependencies
are {\it not} negligible. While
at larger values of $m_\pi$,
where the $\Delta$ is stable, the ratios display a smooth
$m_\pi$ dependence, at $m_\pi =\vDe $ there is an inflection point, and
for  $m_\pi \leq \vDe$ the non-analytic effects are crucial.
The $m_\pi$ dependence obtained here from \ceft\  clearly shows that
the lattice results for $R_{SM}$ may in fact be consistent
with experiment.

\section{Radiative pion photoproduction}

The radiative pion photoproduction ($\gamma N \to \pi N \gamma^\prime$) in the
$\De$-resonance region is used to access
the $\De^+$ magnetic dipole
moment (MDM)~\cite{Machavariani:1999fr,Drechsel:2000um,Drechsel:2001qu}.
The pioneering experiment\cite{Kotulla:2002cg} was carried out at MAMI in 2002 and a series
of dedicated experiments were run in 2005 by the Crystal Ball
Collaboration with first results announced at this school~\cite{CB_Erice07}.
\newline
\indent
In this process, the energy flow can be defined
by the energies of incoming and outgoing photon.
To access the MDM of the $\De$
the energy of the incoming photon must be sufficient
to excite the resonance, while the emitted
photon must be soft. Therefore, in computing
this process, one uses a chiral
expansion with $\De$-isobar degrees of freedom,
the $\de$-expansion, and simultaneously
the soft-photon expansion with respect to
the energy of the emitted photon. In Ref.~\cite{PV_MDM07}, the soft-photon
expansion is performed to the next-next-to-leading order,
since this is the order at which the MDM first appears, while
the chiral expansion is performed to next-to-leading order.
\newline
\indent
An interesting effect which can be studied here is
the absorptive (or, imaginary) part of the MDM~\cite{PV05,Hacker:2006gu}.
It arises due
to the unstable nature of the $\Delta$-isobar. In our \ceft\
calculation, for instance, we find the following result
for the $\De^+$ MDM (in the heavy-baryon limit):
\beq
\mathrm{Im}\, \mu_{\De^{+}}  =  \frac{h_A^2 M_\De}{48\pi f_\pi^2}
\sqrt{\vDe^2 -m_\pi^2}  \,(e/2M_\De)\,.
\eeq
The absorptive MDMs
quantify the change in the width
of the resonance that occurs in an external magnetic field $B$:
\beq
\De \Gamma = 2 \, \mathrm{Im}\, \mu_{\De} \, \vec B\cdot\vec n_s\,,
\eeq
where $\vec n_s$
is the direction of the resonance's spin.
Equivalently, one may look for a change in the
lifetime of the resonance:
$
\De \tau/\tau = -2 \, \mathrm{Im}\, \mu_{\De} \, \vec B\cdot\vec n_s\,\tau,
$
where $\tau = 1/\Gamma$ is the lifetime.
Such a change in the lifetime appears to be extremely small
in moderate magnetic fields and is difficult to be
observed directly~\cite{Binosi:2007ye}.
There is perhaps a possibility to compute the absorptive MDMs
of hadron resonances in lattice QCD
where the effect of arbitrarily large magnetic
fields on the width can  be studied.
\newline
\indent
The \ceft\ description of
the $\gamma p \to \pi^0 p \gamma^\prime$ unpolarized cross section was
found to be consistent with first experimental data for
this process\cite{PV_MDM07}.
It appears that, at low energies of the outgoing
photon, the dependence of the cross-section and linear-photon
asymmetries on the MDM is quadratic, i.e., depends on $|\mu_\De|^2$.
The asymmetry for a circularly
polarized photon beam, however, displays a linear dependence on the
$\Delta^+$ MDM. The helicity difference
for a circularly polarized photon beam
vanishes when approaching the soft-photon limit,
with a rate that is proportional to the MDM. Therefore, a
dedicated measurement with a circularly polarized photon beam could
provides a model-independent extraction of the $\Delta$ MDM.
I refer to the recent paper~\cite{PV_MDM07} for further details.

\section{Summary}

In the single-nucleon sector the limit of applicability of
chiral perturbation theory is set by the excitation energy
of the first nucleon resonance -- the $\De$(1232).
Inclusion of the $\De$ in the chiral Lagrangian extends
the limit of applicability into the resonance energy region.
The power counting of the $\De$ contribution depends crucially
on how the $\vDe=M_\De-M_N$, weighted in comparison
to the other mass scales in the problem, in this case the
pion mass $m_\pi$ and the scale of chiral symmetry breaking $\La$.
\newline
\indent
Two different schemes exist in the literature.
In the Small Scale Expansion, $\vDe \sim m_\pi \ll \La$, while
in the ``$\de$-expansion'', $m_\pi \ll \vDe \ll \La$.
The hierarchy of scales used in the $\de$ expansion
provides a more adequate power-counting
of the $\De$-resonance contributions. It
provides a justification for ``integrating out'' the resonance
contribution at very low energies and for a resummation
and dominance of resonant contributions in the resonance region.
The $\de$ expansion has already been successfully applied
to the calculation of observables for processes
such as Compton scattering,
pion electroproduction and radiative pion photoproduction
in the $\De$-resonance region.

%\section{Acknowledgments}


\begin{thebibliography}{10}
\itemsep -2pt 

\bibitem{Weinberg:1978kz}
  S.~Weinberg,
  %``Phenomenological Lagrangians,''
  Physica A {\bf 96}, 327 (1979);
  %%CITATION = PHYSA,A96,327;%%
%\bibitem{Gasser:1983yg}
  J.~Gasser and H.~Leutwyler,
  Nucl.\ Phys.\ B {\bf 250}, 465 (1985).
  %%CITATION = NUPHA,B250,465;%%

\bibitem{Caprini:2005zr}
  I.~Caprini, G.~Colangelo and H.~Leutwyler,
  %``Mass and width of the lowest resonance in QCD,''
  Phys.\ Rev.\ Lett.\  {\bf 96}, 132001 (2006).
  %[arXiv:hep-ph/0512364].
  %%CITATION = HEP-PH 0512364;%%

\bibitem{vanKolck:2004te}
    P.~F.~Bedaque, H.~W.~Hammer and U.~van Kolck,
  %``Narrow resonances in effective field theory,''
  Phys.\ Lett.\ B {\bf 569}, 159 (2003);
  %[arXiv:nucl-th/0304007].
  %%CITATION = NUCL-TH 0304007;%%
U.~van Kolck,
  %``Effective Field Theories of Light Nuclei,''
  Nucl.\ Phys.\ A {\bf 752}, 145 (2005).
  %[arXiv:nucl-th/0409064].
  %%CITATION = NUCL-TH 0409064;%%



\bibitem{Pascalutsa:2002pi}
  V.~Pascalutsa and D.~R.~Phillips,
  % ``Effective theory of the Delta(1232) in Compton scattering off the
  %nucleon,''
  Phys.\ Rev.\ C {\bf 67}, 055202 (2003).
  %[arXiv:nucl-th/0212024].
  %%CITATION = NUCL-TH 0212024;%%


\bibitem{GSS89}
J.~Gasser, M.~E.~Sainio and A.~Svarc,
%``Nucleons With Chiral Loops,''
Nucl.\ Phys.\ B {\bf 307}, 779 (1988).
%%CITATION = NUPHA,B307,779;%%



\bibitem{Pascalutsa:2006up}
  V.~Pascalutsa, M.~Vanderhaeghen and S.~N.~Yang,
  %``Electromagnetic excitation of the Delta(1232) resonance,''
  Phys.\ Rept.\  {\bf 437}, 125 (2007).
  %[arXiv:hep-ph/0609004].
  %%CITATION = PRPLC,437,125;%%


\bibitem{HHK97}
T.~Hemmert, B.~R.~Holstein and J.~Kambor,
%``Systematic 1/M expansion for spin 3/2 particles in baryon chiral  perturbation theory,''
Phys.\ Lett.\ B {\bf 395}, 89 (1997);
 %%CITATION = HEP-PH 9606456;%%
  %``Chiral Lagrangians and Delta(1232) interactions: Formalism,''
  J.\ Phys.\ G {\bf 24}, 1831 (1998).
  %[arXiv:hep-ph/9712496].
  %%CITATION = HEP-PH 9712496;%%



\bibitem{Gellas:1998wx}
  G.~C.~Gellas {\it et al.},
%T.~R.~Hemmert, C.~N.~Ktorides and G.~I.~Poulis,
  %``The Delta nucleon transition form factors in chiral perturbation  theory,''
  Phys.\ Rev.\ D {\bf 60}, 054022 (1999).
  %[arXiv:hep-ph/9810426].
  %%CITATION = HEP-PH 9810426;%%

\bibitem{Gail}   T.~A.~Gail and T.~R.~Hemmert,
  %``Signatures of chiral dynamics in the nucleon to Delta transition,''
  Eur.\ Phys.\ J.\  A {\bf 28}, 91 (2006).
  %[arXiv:nucl-th/0512082].
  %%CITATION = EPHJA,A28,91;%%

\bibitem{Pascalutsa:2005ts}
  V.~Pascalutsa and M.~Vanderhaeghen,
%``Electromagnetic nucleon to Delta transition in chiral effective-field
  %theory,''
Phys. Rev. Lett. {\bf 95}, 232001 (2005).
%arXiv:hep-ph/0508060.
%%CITATION = HEP-PH 0508060;%%

\bibitem{Pascalutsa:2005vq}
  V.~Pascalutsa and M.~Vanderhaeghen,
  %``Chiral effective-field theory in the Delta(1232) region. I: Pion
%electroproduction on the nucleon,''
  Phys.\ Rev.\ D {\bf 73}, 034003 (2006).
  %[arXiv:hep-ph/0512244].
  %%CITATION = HEP-PH 0512244;%%


\bibitem{Pascalutsa:2005es}
  V.~Pascalutsa, C.~E.~Carlson and M.~Vanderhaeghen,
  % ``Two-photon exchange effects in the electro-excitation of the Delta
  %resonance,''
  Phys.\ Rev.\ Lett.\  {\bf 96}, 012301 (2006).
  %[arXiv:hep-ph/0509055].
  %%CITATION = HEP-PH 0509055;%%

\bibitem{Kondratyuk:2006ig}
  S.~Kondratyuk and P.~G.~Blunden,
  % ``Calculation of two-photon exchange effects for Delta production in
  %electron proton collisions,''
    Nucl.\ Phys.\ A {\bf 778}, 44 (2006).
  %%CITATION = NUCL-TH 0601063;%%


\bibitem{KY99}
S.~S. Kamalov and S.~N. Yang, Phys.\ Rev.\ Lett. {\bf 83}, 4494 (1999);
S.~S. Kamalov, S.~N. Yang, D. Drechsel, O. Hanstein, and L. Tiator,
Phys.\ Rev.\ C {\bf 64}, 032201(R) (2001).

\bibitem{SL}
T. Sato and T.-S.H. Lee,
Phys.\ Rev.\ C {\bf 54}, 2660 (1996);
{\it ibid.}\ {\bf 63}, 055201 (2001).

\bibitem{DUO}
V. Pascalutsa and J.~A. Tjon, 
%``Pion nucleon interaction in a covariant hadron exchange model,''
  Phys.\ Rev.\ C {\bf 61}, 054003 (2000);
  %[arXiv:nucl-th/0003050].
  %%CITATION = NUCL-TH 0003050;%%
{\it ibid.} {\bf 70}, 035209 (2004); 
  %%CITATION = NUCL-TH 0407068;%%
G.~L. Caia {\it et al.}, 
  Phys.\ Rev.\ C {\bf 70}, 032201(R) (2004); 
  %%CITATION = NUCL-TH 0407069;%%
{\it ibid.} {\bf 72}, 035203 (2005).
  %%CITATION = NUCL-TH 0506006;%%

\bibitem{Mertz:1999hp}
  C.~Mertz {\it et al.},
  %``Search for quadrupole strength in the electro-excitation of the
  %Delta(1232)+,''
  Phys.\ Rev.\ Lett.\  {\bf 86}, 2963 (2001).
  %[arXiv:nucl-ex/9902012].
  %%CITATION = NUCL-EX 9902012;%%

\bibitem{Kunz:2003we}
  C.~Kunz {\it et al.},
  %``Measurement of the transverse-longitudinal cross sections in the
  %p(e(pol.),e' p)pi0 reaction in the Delta region,''
  Phys.\ Lett.\ B {\bf 564}, 21 (2003).
%  [arXiv:nucl-ex/0302018].
  %%CITATION = NUCL-EX 0302018;%%



\bibitem{Sparveris:2004jn}
  N.~F.~Sparveris {\it et al.}  [OOPS Collaboration],
  %``Investigation of the conjectured nucleon deformation at low momentum
  %transfer,''
  Phys.\ Rev.\ Lett.\  {\bf 94}, 022003 (2005).
  %[arXiv:nucl-ex/0408003].
  %%CITATION = NUCL-EX 0408003;%%


\bibitem{Pospischil:2000ad}
  T.~Pospischil {\it et al.},
  %``Measurement of the recoil polarization in the p(e(pol.),e' p(pol.))pi0
  %reaction at the Delta(1232) resonance,''
  Phys.\ Rev.\ Lett.\  {\bf 86}, 2959 (2001).
  %[arXiv:nucl-ex/0010020].
  %%CITATION = NUCL-EX 0010020;%%

\bibitem{Alexandrou:2004xn}
  C.~Alexandrou {\it et al.},
  %``The N to Delta electromagnetic transition form factors from lattice QCD,''
  Phys.\ Rev.\ Lett.\  {\bf 94}, 021601 (2005).
  %[arXiv:hep-lat/0409122].
  %%CITATION = HEP-LAT 0409122;%%

\bibitem{Machavariani:1999fr}
  A.~I.~Machavariani, A.~F\" a\ss ler and A.~J.~Buchmann,
  %``Field-theoretical description of electromagnetic Delta resonance
  %production and determination of the magnetic moment of the Delta+(1232)
  %resonance by the e p --> e' N' pi' gamma' and  gamma p --> N' pi' gamma'
  %reactions,''
  Nucl.\ Phys.\  A {\bf 646}, 231 (1999)
  [Erratum-ibid.\  A {\bf 686}, 601 (2001)].

\bibitem{Drechsel:2000um}
  D.~Drechsel {\it et al.}, 
%M.~Vanderhaeghen, M.~M.~Giannini and E.~Santopinto,
  %``Inelastic photon scattering and the magnetic moment of the Delta(1232)
  %resonance,''
  Phys.\ Lett.\ B {\bf 484}, 236 (2000).
  %[arXiv:nucl-th/0003035].
  %%CITATION = NUCL-TH 0003035;%%

\bibitem{Drechsel:2001qu}
  D.~Drechsel and M.~Vanderhaeghen,
% ``Magnetic dipole moment of the Delta(1232)+ from the gamma p --> gamma  pi0
  %p reaction,''
  Phys.\ Rev.\ C {\bf 64}, 065202 (2001).
  %[arXiv:hep-ph/0105060].
  %%CITATION = HEP-PH 0105060;%%

\bibitem{Kotulla:2002cg}
  M.~Kotulla {\it et al.},
% ``The reaction gamma p --> pi0 gamma' p and the magnetic dipole moment of the
  %Delta(1232)+ resonance,''
  Phys.\ Rev.\ Lett.\  {\bf 89}, 272001 (2002).
  %[arXiv:nucl-ex/0210040].
  %%CITATION = NUCL-EX 0210040;%%

\bibitem{CB_Erice07}
M. Kotulla (for the Crystal Ball @ MAMI Coll.), contribution
to these  proceedings.


\bibitem{PV_MDM07}
  V.~Pascalutsa and M.~Vanderhaeghen,
  %``Chiral effective-field theory in the Delta(1232) region: II. radiative pion
  %photoproduction,''
  arXiv:0709.4583 [hep-ph].
  %%CITATION = ARXIV:0709.4583;%%
\bibitem{PV05}
  V.~Pascalutsa and M.~Vanderhaeghen,
  %``Magnetic moment of the Delta(1232)-resonance in chiral effective field
  %theory,''
  Phys.\ Rev.\ Lett.\  {\bf 94}, 102003 (2005).
%  [arXiv:nucl-th/0412113].
  %%CITATION = NUCL-TH 0412113;%%


\bibitem{Hacker:2006gu}
  C.~Hacker, N.~Wies, J.~Gegelia, and S.~Scherer,
  %``Magnetic dipole moment of the Delta(1232) in chiral perturbation theory,''
   Eur.\ Phys.\ J.\  A {\bf 28}, 5 (2006).
  %[arXiv:hep-ph/0603267].
  %%CITATION = EPHJA,A28,5;%%


\bibitem{Binosi:2007ye}
  D.~Binosi and V.~Pascalutsa,
  %``The effect of electromagnetic fields on the lifetime of unstable
  %particles,''
  arXiv:0704.0377 [hep-ph].
  %%CITATION = ARXIV:0704.0377;%%




\end{thebibliography}
\end{document}